\documentclass[12pt,aps,english]{revtex4}
\usepackage{graphicx}
\usepackage{bm}
\usepackage{dcolumn}
\usepackage[usenames, dvipsnames]{color}
\usepackage{epsfig}
\usepackage{amsmath}
\usepackage{mathrsfs}

\begin{document}

\title{Effects of active fluctuations on energetics of a colloidal particle: superdiffusion, dissipation and entropy production}
\author{Subhasish Chaki and Rajarshi Chakrabarti*}
\affiliation{Department of Chemistry, Indian Institute of Technology Bombay, Mumbai, Powai 400076, E-mail: rajarshi@chem.iitb.ac.in}
\date{\today}

\maketitle

\section{Abstract}
\noindent We consider a colloidal particle immersed in an active bath and derive a Smoluchowski equation that governs the dynamics of colloidal particle. We address this as active Smoluchowski equation. Our analysis based on this active Smoluchowski equation shows a short time superdiffusive behavior that strongly depends on the activity. Our model also predicts a non-monotonic dependence of mean energy dissipation against time, a signature of activity-induced dynamics. By introducing a frequency-dependent effective temperature, we show that the mean rate of entropy production is time dependent  unlike a  thermal system. The prime reason for these anomalies is the absence of any fluctuation-dissipation theorem for the active noise. We also comment on how microscopic details of activity can reverse the trends for mean energy dissipation and mean rate of entropy production.

\section{Introduction}

\noindent Active matter systems refer to a class of non-equilibrium systems typically associated with correlated and systematic motion, originating from the imbalance between the energy supplied and the heat dissipated.  Length scales associated with active matter span several orders of magnitude and  include a wide range of phenomena such as the motion of molecular motors \cite{sonn2017scale} and that of chromosomal loci in eukaryotic nuclei \cite{liu2015ghost,zhang2016first}, height undulations of active membranes in red blood cells \cite{park2010measurement}, the movement of bacteria responsible for the  active transport of nutrients in aqueous media \cite{berg2008coli}, search processes \cite{vuijk2018pseudochemotaxis}, active enzymes \cite{jee2018enzyme,jee2018catalytic} and the propulsion of artificial Janus colloids  by diffusiophoretic forces at the microscopic level \cite{howse2007self,io2017experimental} to the macroscale collective dynamics of flocks of birds, schools of fish  \cite{romanczuk2012active}. Systems composed solely of active matters exhibit many fascinating nonequilibrium phenomena such as pattern formation \cite{suzuki2015polar}, self-assembly \cite{angelani2011effective}, phase separation \cite{stenhammar2014phase},  structural organization \cite{needleman2014determining}, periodic beating \cite{chelakkot2014flagellar,sarkar2017spontaneous} whereas in  mixed system (containing a mixture of active and passive particles), structural and dynamical properties of passive particles are surprisingly modified in the presence of active particles. For example Maggi $et\,\,al.$ experimentally and numerically investigated the dynamics of colloidal beads in a bath of swimming $E.coli$ bacteria and found that  collisions from the swimming bacteria lead to enhanced diffusion of the colloid particle \cite{maggi2014generalized}. Recent attempts to model a polymer chain in a bath of bacteria show that the mean square displacement  of a tagged monomer grows faster compare with thermal one and the polymer undergoes swelling \cite{kaiser2014unusual, samanta2016chain, osmanovic2017dynamics, vandebroek2015dynamics, harder2014activity, eisenstecken2017internal, shin2015facilitation, kaiser2015does, osmanovic2018properties, vanderzande2019}.
\\
\\
\noindent  Recently theoretical and experimental attempts have been made  to understand the laws of thermodynamics and non-equilibrium fluctuation relations for active matter \cite{chaudhuri2014active,pietzonka2017entropy,seifert2019stochastic,stuhrmann2012nonequilibrium,fodor2014energetics,krishnamurthy2016micrometre,argun2016non,chaki2018, Goswami2019}. Particularly, the dissipation of energy in active systems can be a useful tool to measure the violation of fluctuation-dissipation theorem (FDT). In this context, Toyabe $et\,\,al.$ have performed single molecule experiments on F1-ATPase motor and quantified the departure from the equilibrium  using Harada-Sasa equality \cite{toyabe2010nonequilibrium}. Experimentally, Bohec $et\,\,al.$ have shown that dissipation of a colloidal bead attached to a specific region of the living cells (such as the cell cortex) can be easily measured using high resolution microscopy, thus enabling one to characterize rheological properties within the cell \cite{bohec2013probing}. However, theoretical as well as experimental studies on  colloidal bead in a bacterial bath to measure active dissipation by force-position correlation is still lacking. In addition, studies based on stochastic thermodynamics for active matter emphasize entropy production as the measure of extracting information from non-equilibrium fluctuations  \cite{seifert2011stochastic,lacoste2009fluctuation,speck2016stochastic,mandal2017entropy,caprini2018comment,dabelow2018irreversibility,dadhichi2018origins, Julicher2017}. A Clausius inequality concerning this has also been proposed \cite{marconi2017,marconi2017heat}. Recently, Fodor $et\,\,al.$ have  modeled non-equilibrium active fluctuations as an Ornstein-Uhlenbeck process \cite{fodor2016far}. They showed that the active system reaches an effective equilibrium and the entropy production rate is zero provided that the persistent time is short. However the probability distribution under such ``effective equilibrium" is not the Boltzmann distribution as associated with systems at thermal equilibrium. The concept of ``effective temperature" has been used widely  as a measure of non-equilibrium fluctuations \cite{samanta2016chain,vandebroek2015dynamics,maggi2014generalized,ben2011effective,berthier2013non} and  has also been realized experimentally \cite{gallet2009}. For nonequilibrium steady state, one can define a time independent effective temperature based on the mean square displacement of the tracer particle \cite{samanta2016chain,vandebroek2015dynamics,flenner2016nonequilibrium,nandi2017nonequilibrium}. However for glassy dynamics, a frequency dependent effective temperature ($T_{eff} (\omega)$) can additionally be defined at any arbitrary time, as the ratio of non-equilibrium spectral density and the response of the system subjected to a small time-dependent perturbation \cite{cugliandolo1997energy}. This idea has also been extended to active systems \cite{201szamel4self,ben2011effective,loi2008effective}. Experimentally Mizuno $et\,\,al.$ have characterized the $T_{eff} (\omega)$ of actomyosin network \cite{mizuno2007nonequilibrium}. In this experiment, the initially cross-linked actin is in thermal equilibrium without myosin. However in the presence of myosin, the network is driven out of equilibrium and  the violation of FDT has been verified using invasive and non-invasive micro-rheology techniques.
\begin{figure*}[h]
\begin{center}
\begin{tabular}{cc}
\includegraphics[width=0.53\textwidth]{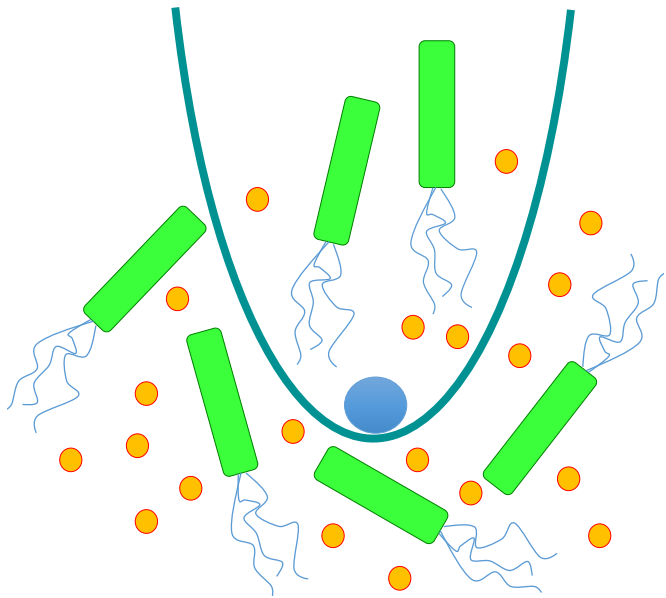} 
\end{tabular}
\end{center}
\caption{Schematic of the model (not to scale): Harmonically trapped single colloidal particle (blue) is immersed in a bath of bacteria (green). The orange balls represent the fluid molecules.}
\label{fig:pic}
\end{figure*}
\\
\\
\noindent Here we aim to investigate the energetics of a tracer bead  immersed in a bath of bacteria  and focus on aspects such as superdiffusion, energy dissipation and entropy production. Our model is  an overdamped, harmonically trapped particle subjected to a thermal and an additional non-equilibrium noise arising from the bacterial bath. The system is initially in thermal equilibrium with the bath. The active noise is modeled as Gaussian random variable, since  the characteristic time scale of the harmonic trap is longer than the correlation time of the active noise \cite{argun2016non}. We proceed by formulating an exact   Smoluchowski equation for the Gaussian active noise.  Our model  ensures superdiffusion at short times as observed in experiments and simulations on active systems \cite{wu2000particle,maggi2014generalized}. Our analysis shows that  the mean energy dissipation from the system to the active bath, measured by force-position correlation has an initial rise which is absent in thermal systems. We also analyze the spectral density of mean energy dissipation rate in our model by Harada-Sasa equality. To understand the deviations from equilibrium, we introduce a $T_{eff} (\omega)$ by applying a small time-dependent perturbation to the particle.  By adopting this notion of $T_{eff} (\omega)$, we investigate mean rate of of entropy production in active bath.
\\
\\
\noindent The paper is organized as follows. In section \ref{Smoluchowski_description} we present a Smoluchowski description of the Gaussian active process and in section \ref{entropy_production} frequency-dependent effective temperature and entropy production are discussed. The energy dissipation in active and thermal bath are presented in section \ref{energy_dissipation} and the paper is concluded in section \ref{conclusion}.

\section{Smoluchowski description of Gaussian active process}  \label{Smoluchowski_description}

\noindent Our model is a harmonically trapped colloidal particle in one dimension subjected to a thermal and a non-equilibrium noise, $\eta_A(t)$ arising from the bacteria bath (Fig. \ref{fig:pic}). The harmonic trap can be attributed to an optical tweezer as used by Argun et.al [33] or to the small amplitude motion of the sedimented colloid at the bottom of the capillary [10].  Active Bio-systems in condensed phases are associated with low Reynolds numbers, so the governing equation of motion for the colloid would be the following overdamped Langevin equation

\begin{equation}
\gamma \frac{dx}{dt} = -kx + \xi_T(t) + \eta_A (t) 
\label{eq:langevinact}
\end{equation}

\noindent Where $\gamma$ is the friction coefficient and $k$ is the spring constant for the harmonic trap and 

\noindent  $\xi_T(t)$ is the Gaussian thermal noise with the statistical properties,

\begin{equation}
\left<\xi_T(t)\right>=0, \left<\xi_T(t) \xi_T(t^\prime)\right> = 2 \gamma k_B T \delta(t-t^\prime)
\end{equation}

\noindent To model active noise, it is essential to note that it originates from the persistent motion of active particles (bacteria). Run-and-Tumble particles, active Brownian particles, and active Ornstein-Uhlenbeck process show such persistent random walk \cite{das2018confined}. However, all three classes exhibit a common noise correlation which decays exponentially with a persistence time $\tau_A$ \cite{zakine2017stochastic,wu2000particle}. In case of  weak trapping and highly viscous medium, the trap relaxation time $\left(\frac{\gamma}{k}\right)$ is longer than the bacterial correlation time $\tau_A$, causing a complete separation of the time scales. This allows us to model the active noise, $\eta_A(t)$ as a Gaussian random variable \cite{argun2016non,vandebroek2017effect, vandebroek2015dynamics, berthier2013non}.  In addition, it has been experimentally shown that the displacement of a tracer bead immersed in an actomyosin network has a Gaussian distribution superimposed with fat exponential tails \cite{stuhrmann2012nonequilibrium, toyota2011non}. For low myosin concentrations the distribution is purely Gaussian \cite{sonn2017scale}. At very low densities of the active particles, it is assumed that these particles interacting weakly. A Gaussian approximation, therefore, works well for these systems \cite{marini2017pressure}. 
\\
\noindent Thus $\eta_A(t)$ is Gaussian and has the following statistical properties
\begin{equation}
\left<\eta_A(t)\right>=0,
\left\langle \eta_A(t) \eta_A(t^{\prime}) \right\rangle=Ce^{-\frac{|t-t^{\prime}|}{\tau_A}}
\label{eq:active_noise_correlation}
\end{equation}

\noindent where $\tau_A$ is the persistence time of the bacterial forces acting on the particle. The prefactor $C$ refers to the  activity which involves the rate of consumption of chemical energy by the bacteria. One should note that the friction in the governing equation of motion for the colloid (Eq.(\ref{eq:langevinact})), has no time dependence, in other words, it is not a generalized Langevin equation \cite{zwanzig2001nonequilibrium}. In support of this assertion, a microrheology experiment on a mixture of passive and active (bacteria) particle has found that the viscosity has no significant frequency dependence \cite{chen2007fluctuations}. A more recent work \cite{maggi2017memory} has also pointed out a clear time scale separation between thermal and active noise, resulting in the breakdown of FDT for active noise. Therefore, to introduce the effects of friction in particle's motion, it is essential to use a white thermal noise, $\xi_T(t)$, as done in our model. However, while modeling chromosomal dynamics or actomyosin network, the inclusion of a frequency dependent friction is essential \cite{zhang2016first,sakaue2017active}, unlike in our case. 
\\
\\
\noindent Initially, the system is in equilibrium with the thermal bath and therefore, the initial position, $x_0$, is chosen from a Boltzmann distribution. Thus we can write $\left<x_0\right>=0$ and $\frac{1}{2}k\left<x_0^2\right>=\frac{1}{2}k_B T$, where $T$ is the temperature of the bath .
\\
\noindent Using Laplace's transformation, we get the solution of Eq.(\ref{eq:langevinact}) for $t>0$ 
\begin{equation}
x(t)=x_0e^{-\frac{k}{\gamma}t}+\frac{1}{\gamma}\int_0^t dt^\prime e^{-\frac{k}{\gamma} (t-t^\prime)}\left(\xi_T(t^\prime) +\eta_A (t^\prime)\right)
\label{eq:langevinact_solution}
\end{equation} 

\noindent  In Eq. (\ref{eq:langevinact_solution}), $x$ is a linear combination of the stochastic variables, $x_0$, $\xi_T$ and $\eta_A$. Therefore, the distribution function of position is Gaussian and the mean and variance is sufficient to find the exact distribution. 
\\
\\
\noindent However, at $t=0$, the system is in thermal equilibrium with the medium, so the initial distribution of the particle is a Boltzmann distribution 
\begin{equation}
P(x_0,0)=\sqrt{\frac{k}{2\pi k_BT}} \exp\left(-\frac{\frac{1}{2}kx_0^2}{k_BT}\right)
\end{equation}

\noindent At any time $t$ $(t > 0)$, the probability distribution is 
\begin{equation}
P(x,t)=\sqrt{\frac{1}{2\pi \left<x^2(t)\right>}} \exp\left(-\frac{x^2(t)}{2\left<x^2(t)\right>}\right)
\end{equation}

\noindent We now introduce the probability distribution, $p(x,t)$ over all possible realizations using the definition \cite{gardinerstochastic}, 
$$p(x,t)=\left<\rho(x,t)\right>=<\delta(x(t)-x)>$$

\noindent The probability current, $J(x,t)$, is defined as $J(x,t)=\rho(x,t)\dot{x}(t)$

\noindent The above continuity equation is given by 
\begin{equation}
\frac{\partial \rho(x,t)}{\partial t}=-\frac{\partial J(x,t)}{\partial x}
\label{eq:continuity}
\end{equation}

\noindent From continuity equation (\ref{eq:continuity}), one obtains the following evolution equation,

\begin{equation}
\begin{split}
 \frac{\partial p(x,t)}{\partial t}&=\frac{k}{\gamma}\frac{\partial}{\partial x}xp(x,t)+\left(\frac{k_B T}{\gamma}+\frac{A(t)}{\gamma}\right) \frac{\partial^2}{\partial x^2}p(x,t)\\
&=\frac{k}{\gamma}\frac{\partial}{\partial x}xp(x,t)+\left(D+D_A(t)\right) \frac{\partial^2}{\partial x^2}p(x,t)
\label{eq:active_fokker}
\end{split}
\end{equation} 
\\
\noindent where the thermal diffusion coefficient is given by $D=\frac{k_B T}{\gamma}$ and we have introduced a new variable here, $D_A(t)=\frac{A(t)}{\gamma}$. One should notice that the time dependence of $D_A (t)$ results from the non-Markovian nature of the active noise. The readers are refereed to Appendix (\ref{appendix_fokker_active}) for a detailed calculations.
\\
\\
In the long time limit, the persistence in the active noise correlation is negligibly small and the system reaches a non-equilibrium steady state with a renormalized diffusivity $D_{\text{renormalized}}=D+D_A$ {\cite{willareth2017generalized}, which is consistent with earlier studies \cite{maggi2014generalized,samanta2016chain,vandebroek2015dynamics,ghosh2014dynamics}, where $D_A=\frac{k_B C}{\gamma^2 \left(\frac{k}{\gamma}+\frac{1}{\tau_A}\right)}=\frac{k_B T_A}{\gamma}$. The long time diffusivity, $D_A$ matches with Eq. (20) of Szamel and for the free diffusion case $(k=0)$ it matches with Eq. (6) of Szamel \cite{201szamel4self}. For the purely thermal case, $C=0$, one gets the well known Smoluchowski equation for a system in thermal equilibrium \cite{zwanzig2001nonequilibrium}. We shall henceforth refer to Eq. (\ref{eq:active_fokker}) as the active Smoluchowski equation (ASE).  
\\
\\
\noindent Gaussian nature of the process ensures that the solution of the above ASE is finite, so as the moments.  In other words $p(x,t)$  decays ``fast'' at long time (specifically, faster than $|x|^{n-1}$ if $n^{th}$ moments  are finite), and so does its first derivative ($\frac{\partial}{\partial x}p(x,t)$) .  To calculate the mean square fluctuation of position, $\left<x^2(t)\right>$ from the ASE, we multiply both sides of Eq.  (\ref{eq:active_fokker}) by $x^2$ and  integrate over all possible values of $x$ to get :

\begin{equation}
\begin{split}
\sigma^2(t)\equiv\left<\left(x(t)-x(0)\right)^2\right>&=\frac{2k_B T}{k}\left(1- e^{-\frac{k}{\gamma} t}\right)+\frac{C}{k\gamma (\frac{k}{\gamma}+\frac{1}{\tau_A})}\left(1-e^{-\frac{2k}{\gamma}t}\right)\\
&-\frac{2C}{(\frac{k^2}{\gamma^2}-\frac{1}{\tau_A^2})\gamma^2}\left(e^{-(\frac{k}{\gamma}+\frac{1}{\tau_A})t}-e^{-\frac{2k}{\gamma}t}\right)
\label{eq:msd_active}
\end{split}
\end{equation} 

\noindent For both $C=0$ or $\tau_A=0$, the system behaves as purely thermal system.

\noindent The readers are refereed to Appendix (\ref{MSD_appendix}) for a detailed calculations.

\noindent With the condition, $\frac{\gamma}{k} > \tau_A$, we can approximate Eq. (\ref{eq:msd_active}) for short time $(t<\frac{\gamma}{k})$, 
\begin{equation}
\begin{split}
\sigma^2(t)&=2Dt+\frac{C t^2}{ \gamma^2}
\label{eq:active_superdiffusion}
\end{split}
\end{equation}
\\
\noindent For $\frac{\gamma}{k}>t>\frac{2D \gamma^2 }{C}$, superdiffusion $(\sigma^2(t)\approx \frac{C t^2}{ \gamma^2})$ is observed whereas thermal diffusion $(\sigma^2(t)=2Dt)$ is observed for $\frac{\gamma}{k}>t<\frac{2D \gamma^2 }{C}$. This is certainly not because of inertia as the governing equation (\ref{eq:langevinact}) is overdamped. With increasing $C$, superdiffusion sets in at an earlier time.
\noindent The readers are refereed to Appendix (\ref{superdiffusion_appendix}) for a detailed calculations.
\\
\begin{figure*}[ht]
\begin{center}
\begin{tabular}{cc}
\includegraphics[width=0.55\textwidth]{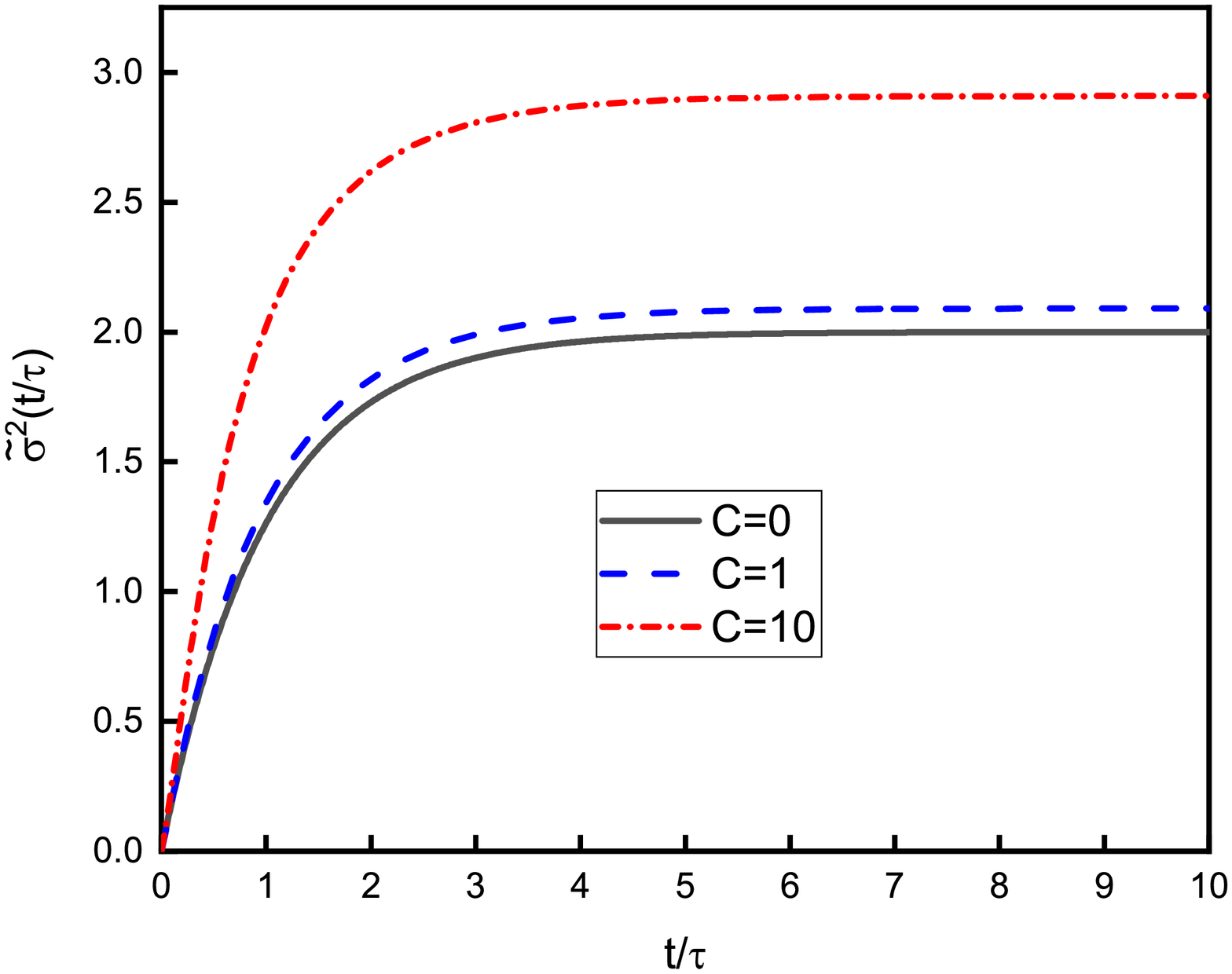} &
\includegraphics[width=0.62\textwidth]{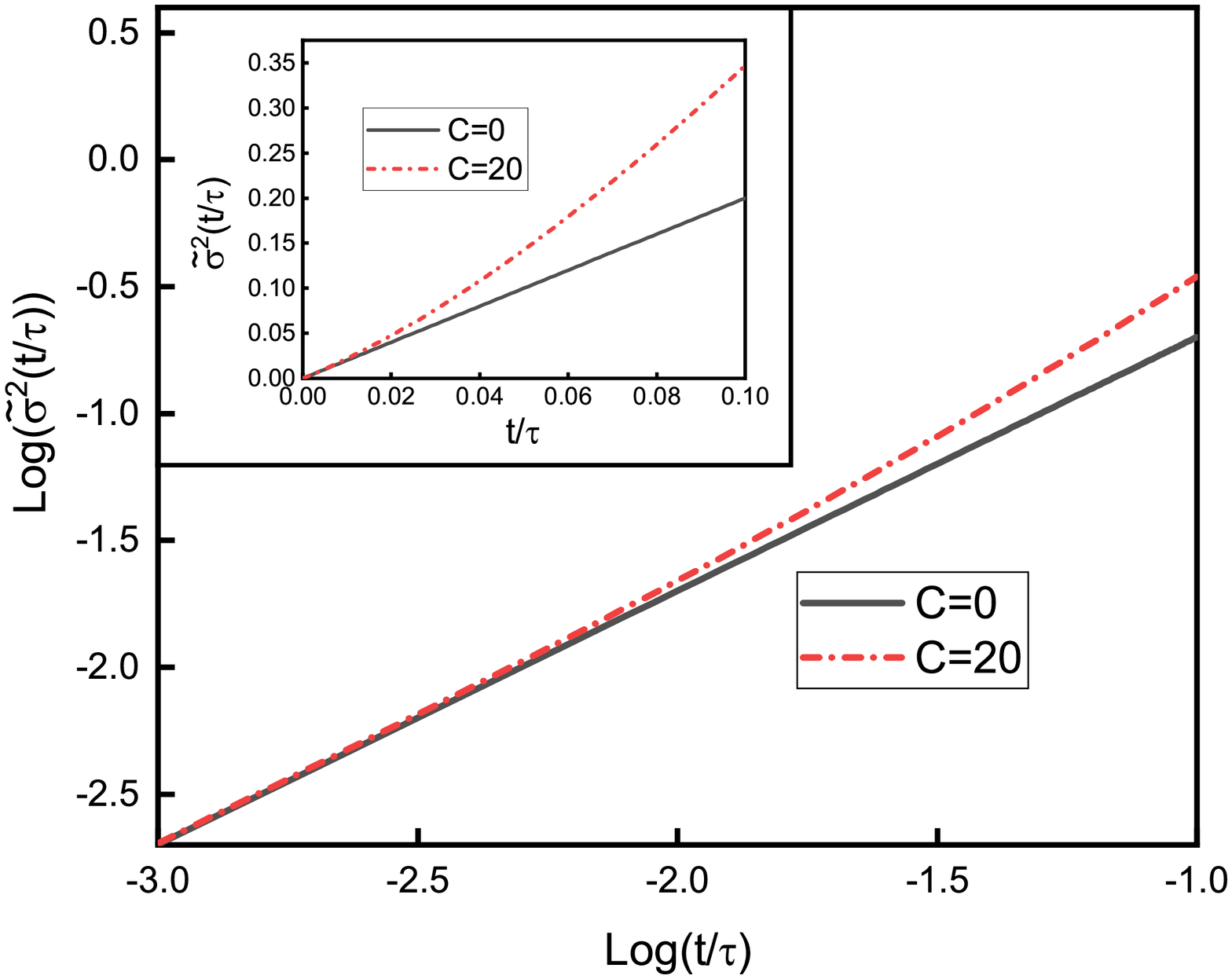}\\
(a) & (b)  \\
\end{tabular}
\end{center}
\caption{Plot of $\tilde{\sigma}^2(t/\tau)$ against $t/\tau$ for single particle in active bath for different values of $C$ (plot (a)).  The values of the parameters used for  (a) are   $\gamma=1, k_B=1, T=1, \tau_A=0.1, k=1$. Log-Log plot of $\tilde{\sigma}^2(t/\tau)$ against $t/\tau$ for single particle in active bath (plot (b)). The values of the parameters used for  (b) are   $k=0.001, \gamma=1, k_B=1, T=1, \tau_A=0.1$.}
\label{fig:active_msd}
\end{figure*}
\\
\noindent Variation of  non-dimensional MSD $\left(\tilde{\sigma}^2(t/\tau_A)=\frac{\sigma^2(t/\tau_A)}{\left<x_0^2\right>}\right)$ against non-dimensional time $(t/\tau_A)$ can be seen in Fig.  (\ref{fig:active_msd} (a)). With increasing $C$, MSD of the particle grows faster. Although we do not consider any non-Gaussianity in the distribution of active noise, we observe superdiffusion with an exponent $\alpha=2$ at short time (when activity is high) as can be seen from Eq. (\ref{eq:active_superdiffusion}) and Fig. (\ref{fig:active_msd} (b)). This is a consequence of the persistent motion of the active particles.
\\
\\
\noindent Here we would also like to comment that the governing equation of motion for the colloid (Eq.(\ref{eq:langevinact})) has a nonequilibrium noise  with  finite temporal correlation and thus the dynamics is non-Markovian. However, if one draws $\eta_A(t)$ from an Ornstein-Uhlenbeck process (OUP) then the governing equation of $\eta_A(t)$  is
\begin{equation}
\tau_A\frac{d\eta_A}{dt} = -\eta_A(t)+  \Gamma(t)
\label{eq:ornstein}
\end{equation}
\noindent With $\left<\Gamma(t)\right>=0$ and $\left\langle \Gamma(t) \Gamma(t^{\prime}) \right\rangle=2C_0\delta(t-t^\prime)$, the following correlation results
\begin{equation}
\left\langle \eta_A(t) \eta_A(t^{\prime}) \right\rangle=\frac{C_0}{\tau_A}e^{-\frac{|t-t^{\prime}|}{\tau_A}}
\label{eq:ou_correlation}
\end{equation}
\noindent Where $C_0$ is the strength of the active noise for OUP. Eq.(\ref{eq:langevinact}) and Eq.(\ref{eq:ornstein}) together make a commensurate system that obeys fluctuation-response theorem (FRT) as the dynamics is Markovian which allows this generalization of FDT in an expanded space of position $(x)$ and active noise $(\eta_A(t))$. But our model does not follow FDT neither it follows FRT for our choice of active noise as it is not drawn from an OUP  \cite{prost2009generalized}. 
\\
\noindent By using Eq.(\ref{eq:ou_correlation}), MSD of the particle for OUP of $\eta_A(t)$ 
\\
\begin{equation}
\begin{split}
\sigma_{\text{OU}}^2(t)&=\frac{2k_B T}{k}\left(1- e^{-\frac{k}{\gamma} t}\right)+\frac{C_0}{k\gamma\tau_A (\frac{k}{\gamma}+\frac{1}{\tau_A})}\left(1-e^{-\frac{2k}{\gamma}t}\right)\\
&-\frac{2C_0}{(\frac{k^2}{\gamma^2}-\frac{1}{\tau_A^2})\gamma^2\tau_A}\left(e^{-(\frac{k}{\gamma}+\frac{1}{\tau_A})t}-e^{-\frac{2k}{\gamma}t}\right)
\label{eq:msd_active_oup}
\end{split}
\end{equation} 
\\
\noindent In the limit $\tau_A\rightarrow 0$, $\sigma_{\text{OU}}^2(t)$ diverges. However, taking this limit at the very beginning, one can show that $\lim_{\tau_A \to 0}\frac{C_0}{\tau_A}e^{-\frac{|t-t^{\prime}|}{\tau_A}}=C_0\delta(t-t^{\prime})$. This ensures that the active noise has no memory and is $\delta$-correlated in time. In other words, it is equivalent of rescaling the temperature. In this case, $\sigma_{\text{OU}}^2(t)$ does not show any divergence in the limit $\tau_A \rightarrow 0$ but shows  normal diffusion with a modified diffusion coefficient instead of superdiffusion. Also, $D_A(t)$ in Eq.(\ref{eq:active_fokker}) will be a constant for the OUP of the active noise as the dynamics becomes Markovian in the expanded space \cite{201szamel4self, nandi2018random}.

\section{Frequency-dependent effective temperature  and  entropy production} \label{entropy_production} 

\noindent In a typical invasive, also known as  active micro-rheology experiment (AMR), a known force $F(t)$ is applied on the tracer bead (or probe particle) and the response is measured from the tracer's displacement whereas in passive or non-invasive micro-rheology experiments (PMR), the spontaneous displacement fluctuations of a tracer bead is measured without applying any external force. At equilibrium, the power spectral density of the displacement fluctuations is directly related to the dissipative part of the response function (imaginary part of response function) as
\\
\begin{equation}
\begin{split}
 \textrm{Im}[\chi(\omega)]&=\frac{\omega}{2k_B T}\textrm{Re}[S(\omega)]\\
\end{split}
\label{eq:FDT}
\end{equation}
\\
\noindent where T is the ambient temperature, $\textrm{Re}[S(\omega)]$ is the real part of the one-sided Fourier transform of position autocorrelation function and $ \textrm{Im}[\chi(\omega)]$ is the imaginary part of the one-sided Fourier transform of the response function. Eq. (\ref{eq:FDT}) is  the FDT. Activity create additional fluctuations and make the right hand side of Eq. (\ref{eq:FDT}) larger than the left hand side, thus violating FDT.
Because we can independently measure the left hand side of Eq. (\ref{eq:FDT}) with AMR and right hand side with PMR, Eq. (\ref{eq:FDT}) serves as a useful tool to characterize non-equilibrium fluctuations \cite{rupprecht2016fresh,mizuno2007nonequilibrium,gomez2009experimental} .
\\
\\
\noindent To calculate the response function, $\chi(t)$, we apply a weak time-dependent force, $F(t)$ on the particle. In this case, the dynamics of the particle is governed by the overdamped Langevin equation,
\begin{equation}
\gamma\frac{dx}{d{t}}=-k x(t)+\xi_T(t)+\eta_A(t)+F(t)
\end{equation}
\\
\noindent Here we assume that both the bath properties are not affected by external driving. We can write the linear response in the form, $\left<\delta x(t)\right>=\int dt^{\prime}\chi(t-t^{\prime})F(t^{\prime})$ where $\left<\delta x(t)\right>=\left<x(t)\right>-\left<x_0\right>$. Since $\left<\xi_T(t)\right>=\left<\eta_A(t)\right>=\left<x_0\right>=0$, we get $\left<\delta x(t)\right>=\frac{1}{\gamma}\int_{0}^{t}dt^{\prime}e^{-\frac{k}{\gamma}(t-t^{\prime})}F(t^{\prime})$.
\\
\noindent Thus
\begin{equation}
\chi(t)=\begin{cases}\frac{1}{\gamma} e^{-\frac{k}{\gamma} t} \,\,\,\,\,t>0 \\ 0 \,\,\,\,\,\,\,\,\,\,\,\,\,\,\,\,\,t<0
\end{cases}
\end{equation}
\\
\noindent This response function, $\chi(t)$ is independent of any form of active drive \cite{ahmed2015active,ahmed2018active}. 
\\
\\
\noindent  The Fourier transformation of the response function is given by
\begin{equation}
\chi(\omega)=\frac{1}{\gamma}\int_{-\infty}^{\infty}dt e^{-\frac{k}{\gamma} t}e^{i\omega t}=\frac{1}{\gamma}\int_{0}^{\infty}dt e^{-\frac{k}{\gamma} t}e^{i\omega t}=\frac{k+i\omega\gamma}{k^2+\omega^2\gamma^2}
\end{equation}

\noindent On the other hand, the definition of  spectral density, $S(\omega)$ is given by $S(\omega)=|\tilde{x}(\omega)|^2$. Thus, to calculate $S(\omega)$, we begin by Fourier transforming  Eq. (\ref{eq:langevinact})

\begin{equation}
\begin{split}
-i\omega \gamma \tilde{x}(\omega)=-k\tilde{x}+\tilde{\xi_T}(\omega)+\tilde{\eta_A}(\omega)\\
\label{eq:fourier_langevin}
\end{split}
\end{equation}

\noindent where $\tilde{x}(\omega), \tilde{\xi_T}(\omega), \tilde{\eta_A}(\omega) $ denote the Fourier transforms of $x,\xi_T,\eta_A$. From Eq.  (\ref{eq:fourier_langevin}) we get,

\begin{equation}
\begin{split}
 \tilde{x}=\frac{\tilde{\xi_T}(\omega)+\tilde{\eta_A}(\omega)}{k-i\omega \gamma}\\
\label{eq:fourier_langevin11}
\end{split}
\end{equation}

\noindent Therefore,
\begin{equation}
\begin{split}
S(\omega)\equiv\left< \tilde{x}(\omega) \tilde{x^{\ast}}(\omega)\right>=\frac{\left< \tilde{\xi_T^2 }(\omega)\right>+\left<\tilde{\eta_A^2}(\omega)\right>}{k^2+\omega^2\gamma^2}\\
\label{eq:fourier_langevin11}
\end{split}
\end{equation}

\noindent where $\left< \tilde{\xi_T^2 }(\omega)\right>$ and $\left<\tilde{\eta_A^2}(\omega)\right>$ are Fourier transformation of the autocorrelation function of $\xi_T$ and $\eta_A$ respectively (Wiener-Khinchin theorem \cite{balakrishnan2008elements}) and   $\left< \tilde{\xi_T^2 }(\omega)\right>=2\gamma k_B T$ and   $\left<\tilde{\eta_A^2}(\omega)\right>=\frac{C}{\frac{1}{\tau_A}-i\omega}=\frac{C\tau_A}{1-i\omega\tau_A}$.  

\begin{figure*}[h]
\begin{center}
\begin{tabular}{cc}
\includegraphics[width=0.57\textwidth]{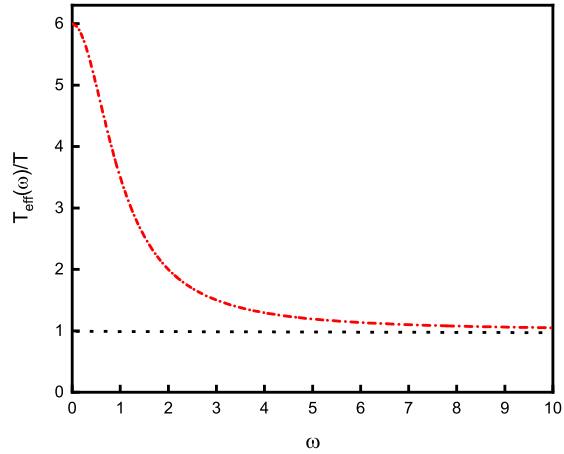} 
\end{tabular}
\end{center}
\caption{Plot of $T_{\text{eff}} (\omega)/T$ against $\omega$ for single particle in active bath. The values of the parameters used are $\gamma=1, k_B=1, T=1, \tau_A=1, C=1$.}
\label{fig:T_eff_omega}
\end{figure*}

\noindent By substituing $\textrm{Re}[S(\omega)]$ and $\textrm{Im}[\chi(\omega)]$ in Eq. (\ref{eq:FDT}), one can define $T_{\text{eff}}(\omega)$ for active system as

\begin{equation}
\begin{split}
T_{\text{eff}}(\omega)&=T+\frac{C\tau_A}{2k_B \gamma \left(1+\omega^2\tau_A^2\right)}\\
&=\frac{2T \left(1+\omega^2\tau_A^2\right)+T_{A,k=0}}{2 \left(1+\omega^2\tau_A^2\right)}
\label{eq:effective_temp_freq}
\end{split}
\end{equation}
\noindent where, $T_{A,k=0}$ is the active temperature for free diffusion \cite{wu2000particle}. In the high-frequency limit, $T_{\text{eff}}(\omega)$ is dominated by the thermal noise, so that $T_{\text{eff}}(\omega)\rightarrow T$. In the small frequency limit $(\omega \rightarrow 0)$, $T_{\text{eff}}(\omega)$ approaches a constant value : $T+\frac{C\tau_A}{2k_B \gamma}$. 
\\
\noindent The inverse Fourier transformation of $\frac{1}{T_{\text{eff}}(\omega)}$,
\\
\begin{equation}
\begin{split}
T_{\text{eff}}^{-1}(t)&=\frac{\delta(t)}{T} - \frac{T_{A,k=0}\sqrt{\frac{T\tau_A^2}{2T+T_{A,k=0}}}\exp \Bigg[{-\frac{t}{\sqrt{\frac{2T\tau_A^2}{2T+T_{A,k=0}}}} \Bigg]}}{2T^2\tau_A^2}\\
&=\frac{\delta(t)}{T} - \frac{T_{A,k=0}\sqrt{\frac{T}{2T+T_{A,k=0}}}\exp \Bigg[{-\frac{t}{\tau_A\sqrt{\frac{2T}{2T+T_{A,k=0}}}} \Bigg]}}{2T^2\tau_A}\\
\label{eq:effective_temp_time}
\end{split}
\end{equation}
\\
\noindent For $C=0$, the bath is in thermal equilibrium at temperature $T$ and thus, $T_{\text{eff}}^{-1}(t)=\frac{\delta(t)}{T}$. 
\\
\noindent In stochastic thermodynamics, physical quantities such as internal energy $\Delta U$, dissipated heat $Q$ and Jarzynski's work $W_J$ are related as  \cite{ritort2007nonequilibrium}
\\
\begin{equation}
Q=W_J-\Delta U
\label{eq:first_law}
\end{equation}
\\
\noindent Equation (\ref{eq:first_law})  is a statement of the first law of thermodynamics. In our case, we want to investigate the case of entropy production in the presence of active fluctuations keeping aside the effect of weak time-dependent force $F(t)$ and therefore, $W_J \cong 0$ and  $\Delta U=\frac{1}{2}kx^2 - \frac{1}{2}kx_0^2$.
\\
\noindent The change of entropy in the medium over the time interval $t$  \cite{zamponi2005fluctuation},
\\
\begin{equation}
\Delta S_m = -\int_0^t dt^\prime \left[\frac{1}{2}kx^2(t^\prime)-\frac{1}{2}kx_0^2\right]T_{eff}^{-1}(t^\prime)
\end{equation}
\\
\noindent The definition of non-equilibrium Gibbs entropy $S(t)=-k_B \int dx p(x,t) \ln p(x,t)$ leads to the definition of a trajectory-dependent entropy in the system $s(t)=-k_B \ln p(x,t)$ such that $S(t)=\left<s(t)\right>$ \cite{seifert2005entropy}.  
\\
\noindent Thus the total change in entropy is
\begin{equation}
\begin{split}
\Delta S_{tot}&=\Delta S_m + \Delta s  \\
&= -\int_0^t dt^\prime \left[\frac{1}{2}kx^2(t^\prime)-\frac{1}{2}kx_0^2\right]T_{\text{eff}}^{-1}(t^\prime)-k_B \ln p(x,t)+k_B \ln p(x_0,0)
\end{split}
\end{equation}
\\
\noindent We would like to comment that if one uses $\delta(t-t^\prime)T^{-1}$ instead of $T_{eff}^{-1}(t^\prime)$ to calculate the entropy change for the medium, earlier results by Chaki and Chakrabarti \cite{chaki2018} is recovered. This would be the change of entropy of the medium due to heat exchange with the thermal environment of constant ambient temperature $T$. 
\\
\noindent Using ASE (Eq. (\ref{eq:active_fokker})), the rate of change of total entropy follows as, 
\\
\begin{equation}
\begin{split}
\frac{d \Delta S_{tot}}{dt}&= - \left[\frac{1}{2}kx^2(t)-\frac{1}{2}kx_0^2\right]T_{eff}^{-1}(t)- \frac{k_B \partial_t p(x,t)}{p(x,t)}
\label{eq:rate_entropy}
\end{split}
\end{equation}
\\
\begin{figure*}[h]
\begin{center}
\begin{tabular}{cc}
\includegraphics[width=0.57\textwidth]{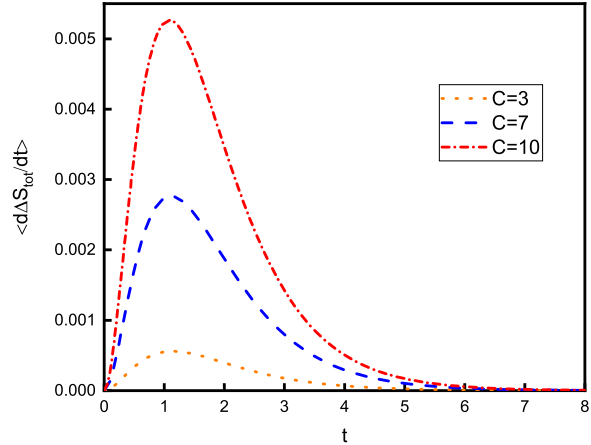} 
\end{tabular}
\end{center}
\caption{Plot of $\left<\frac{d \Delta S_{tot}}{dt}\right>$ against $t$ for single particle in active bath for different values of $C$. The values of the parameters used are   $k=1, \gamma=2, \tau_A=1$.}
\label{fig:rate_entropy}
\end{figure*}
\\
\noindent Eq. (\ref{eq:rate_entropy}) represents   $\frac{d \Delta S_{tot}}{dt}$, as a fluctuating quantity because it depends on stochastic variable $x(t)$. Therefore, it is convenient to expressed Eq. (\ref{eq:rate_entropy}) by its averaged quantity, the mean rate of change of total entropy over many realizations,
\\
\begin{equation}
\begin{split}
\left<\frac{d \Delta S_{tot}}{dt}\right>&= - \left[\frac{1}{2}k\left<x^2(t)\right>-\frac{1}{2}k\left<x_0^2\right>\right]T_{eff}^{-1}(t)- k_B \int dx \frac{ \partial_t p(x,t)}{p(x,t)} p(x,t)\\
&=-\left[\frac{k_B T_A}{2}\left(1-e^{-\frac{2k}{\gamma}t}\right)-\frac{C k\left(e^{-\left(\frac{k}{\gamma}+\frac{1}{\tau_A}\right)t}-e^{-\frac{2k}{\gamma}t}\right)}{\gamma^2\left(\frac{k^2}{\gamma^2}-\frac{1}{\tau_A^2}\right)} \right]\\
& \times \left[\frac{\delta(t)}{T} - \frac{T_{A,k=0}\sqrt{\frac{T}{2T+T_{A,k=0}}}\exp \Bigg[{-\frac{t}{\tau_A\sqrt{\frac{2T}{2T+T_{A,k=0}}}} \Bigg]}}{2T^2\tau_A}\right] - k_B \partial_t \int dx p(x,t)\\
&=-\left[\frac{k_B T_A}{2}\left(1-e^{-\frac{2k}{\gamma}t}\right)-\frac{C k\left(e^{-\left(\frac{k}{\gamma}+\frac{1}{\tau_A}\right)t}-e^{-\frac{2k}{\gamma}t}\right)}{\gamma^2\left(\frac{k^2}{\gamma^2}-\frac{1}{\tau_A^2}\right)} \right]\\
& \times \left[\frac{\delta(t)}{T} - \frac{T_{A,k=0}\sqrt{\frac{T}{2T+T_{A,k=0}}}\exp \Bigg[{-\frac{t}{\tau_A\sqrt{\frac{2T}{2T+T_{A,k=0}}}} \Bigg]}}{2T^2\tau_A}\right]
\end{split}
\end{equation}
\\
\noindent Here we have used $\frac{\partial}{\partial t}\int dx p(x,t)=0$. However, conventionally in stochastic thermodynamics,  $\left<\frac{d \Delta S_{tot}}{dt}\right>$ is interpreted as the mean rate of entropy production \cite{zamponi2005fluctuation}. 
\\
\noindent In Fig. (\ref{fig:rate_entropy}), for small values of time, mean rate of change of total entropy, $\left<\frac{d \Delta S_{tot}}{dt}\right>$ increases gradually, reflecting no FDT in active system whereas, for thermal system $\left<\frac{d \Delta S_{tot}}{dt}\right>=0$. This time dependence of $\left<\frac{d \Delta S_{tot}}{dt}\right>$ at short time, leads to the breakdown of time-reversal symmetry which is a signature of nonequilibrium induced by activity. 

\section{Energy dissipation in active bath} \label{energy_dissipation} 
\begin{figure*}[h]
\begin{center}
\begin{tabular}{cc}
\includegraphics[width=0.57\textwidth]{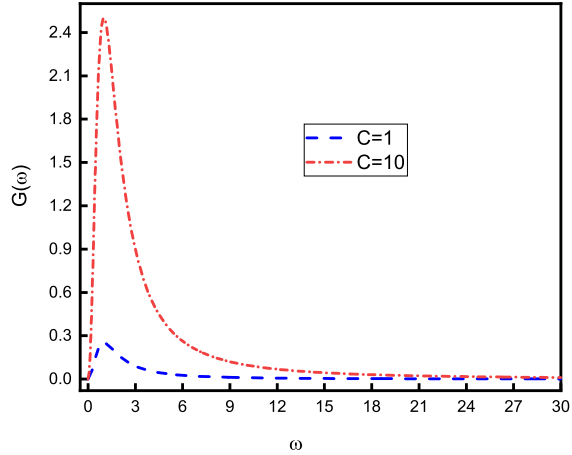} 
\end{tabular}
\end{center}
\caption{Plot of $G(\omega)$ against $\omega$ for single particle in active bath for different values of $C$. The values of the parameters used are   $k=1, \gamma=1, \tau_A=1$.}
\label{fig:dissipation_rate}
\end{figure*}

\noindent Recently, a nonequilibrium equality was derived by Harada and Sasa which connects the response function and the correlation function \cite{dadhichi2018origins,harada2005equality}: 
\begin{equation}
J=\frac{\gamma}{2\pi}\int_{-\infty}^{\infty} d\omega  \left[\omega S(\omega) - 2k_B T \text{Im}\chi(\omega)\right]\omega
\label{eq:sasa_dissipation}
\end{equation}
where $J$ is referred as the mean rate of energy dissipation. Since $J=0$ at equilibrium, any finite $J$ can be a measure of deviation from equilibrium. Substituing $S(\omega)$ and $\text{Im}\chi(\omega)$ in Eq. (\ref{eq:sasa_dissipation}) and then integrating over all possible $\omega$ gives $J=\frac{C}{2k(\tau+\tau_A)}$. This non-zero value of $J$ arises solely because of the presence of activity in the medium. The spectral density of the energy dissipation rate is defined as $G(\omega)=\gamma \left[\omega S(\omega) - 2k_B T \text{Im}\chi(\omega)\right]\omega$. For our model, $G(\omega)=\frac{\gamma C\tau_A \omega^2}{\left(1+\omega^2\tau_A^2 \right) \left(k^2+\omega^2 \gamma^2 \right)}$ and we have plotted $G(\omega)$ against $\omega$ for different values of $C$ in Fig. (\ref{fig:dissipation_rate}).
\\
\\
\noindent  Experimentally, Bohec $et\,\,al.$ have shown that the power spent by the random active forces on the bead can be measured from the time derivative $\frac{d}{dt}\left<x(t)\eta_A(0)\right>$. Thus dissipated energy is related to the cross-correlation between random force exerted on the bead at $t=0$ and its position for the overdamped motion at $t=t$ \cite{bohec2013probing}. Previous studies on active systems, however, focused only on force-force or position-position correlation {\cite{samanta2016chain,osmanovic2017dynamics,vandebroek2015dynamics}. In general, active forces originate from nonequilibrium fluctuations and are independent of ambient temperature $T$. Thus dissipation from the active heat bath can be de-coupled  from the dissipation to the thermal bath, and the total force-position correlation can be written as $\text{I}(t)=\left< x(t)\xi_T(0) \right>+\left< x(t)\eta_A(0) \right>$.
\\
\begin{figure*}[ht]
\begin{center}
\begin{tabular}{cc}
\includegraphics[width=0.55\textwidth]{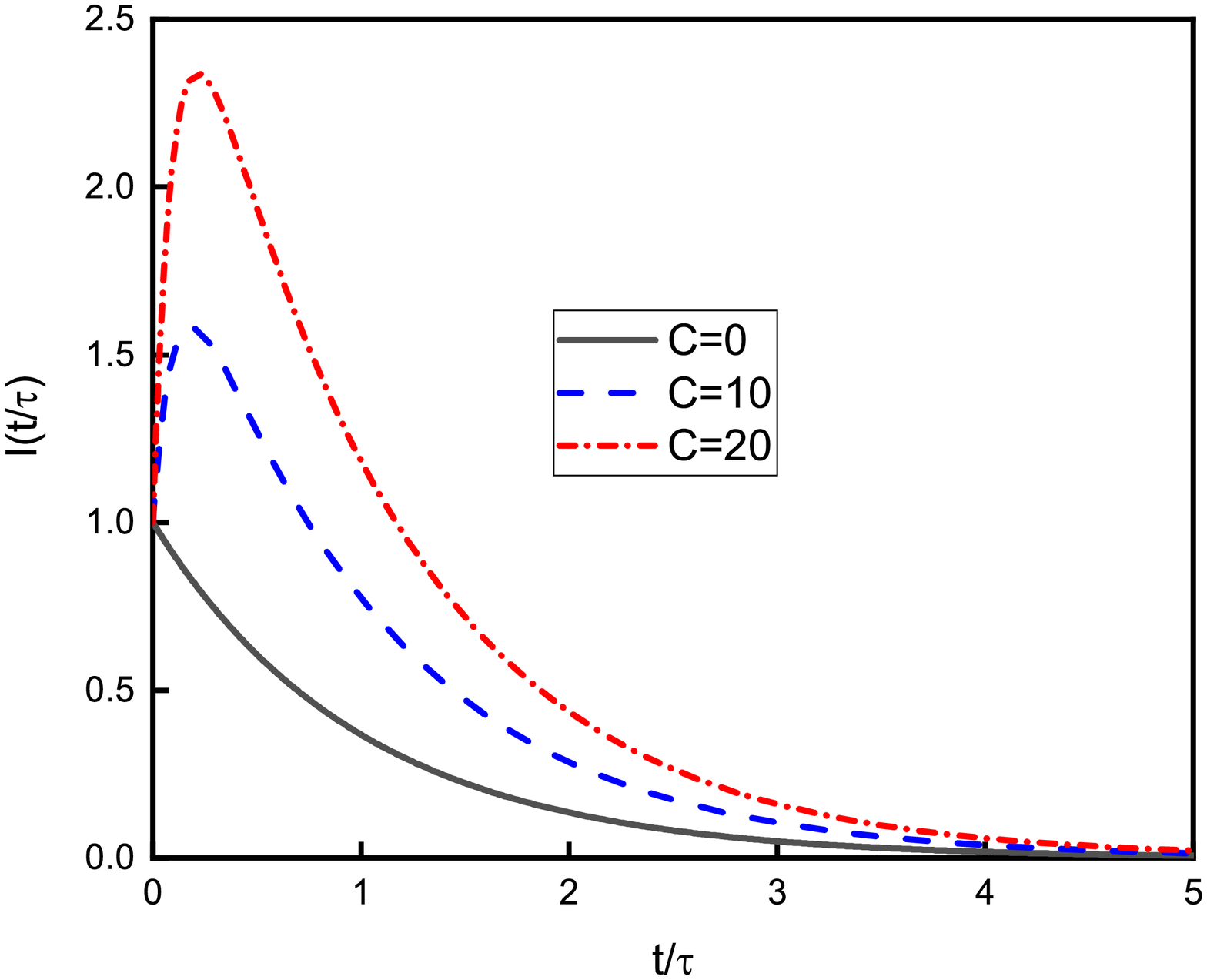} &
\includegraphics[width=0.55\textwidth]{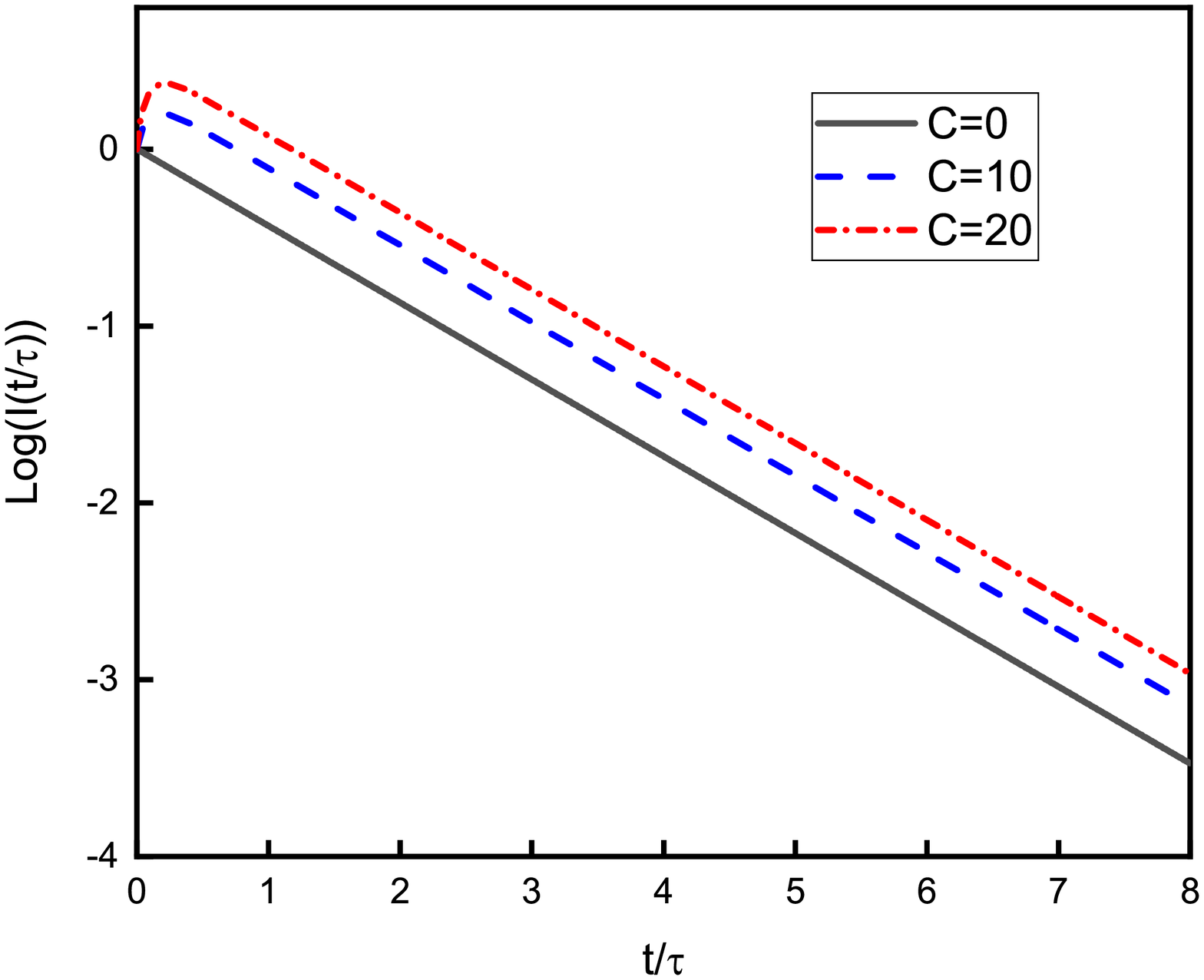}\\
(a) & (b)  \\
\end{tabular}
\end{center}
\caption{(a) Plot of $\text{I}(t/\tau)$ against $t/\tau$ for single particle in active bath (b)  Log-linear Plot of $\text{I}(t/\tau)$ against $t/\tau$. The values of the parameters used are   $k=1, \gamma=1, k_B=1, T=1, \tau_A=0.1$.}
\label{fig:total_dissipation}
\end{figure*}

\noindent  Force-position correlation in an active bath is given by,

\begin{equation}
\begin{split}
\left< x(t)\eta_A(0) \right>&=\frac{C\left[e^{-\frac{t}{\tau_A}}-e^{-\frac{k}{\gamma}t}\right]}{\gamma\left(\frac{k}{\gamma}-\frac{1}{\tau_A}\right)}
\label{eq:active_pos_force_correlation}
\end{split}
\end{equation}

\noindent and the force-position correlation in thermal bath is 

\begin{equation}
\begin{split}
\left< x(t)\xi_T(0) \right> & =k_B T  e^{-\frac{k}{\gamma}t}
\label{eq:thermal_pos_force_correlation}
\end{split}
\end{equation}

\noindent Therefore,  

\begin{equation}
\begin{split}
\text{I}(t) &= k_B T  e^{-\frac{k}{\gamma}t}+\frac{C\left[e^{-\frac{t}{\tau_A}}-e^{-\frac{k}{\gamma}t}\right]}{\gamma\left(\frac{k}{\gamma}-\frac{1}{\tau_A}\right)}
\label{eq:total_pos_force_correlation}
\end{split}
\end{equation}
\noindent The readers are referred to Appendices (\ref{active_dissipation_appendix}) and (\ref{thermal_dissipation_appendix}) for detailed calculations.

\noindent The initial increase of $\text{I}(t/\tau)$ for non-zero $C$, results from a competition between the two time scales: i.e., $\tau_A$, the persistence time of the active noise and $\frac{\gamma}{k}$, the relaxation time in the harmonic well. The initial growth of $\text{I}(t/\tau)$ is magnified with higher $C$ (or $\tau_A$) . On the other hand for thermal system $(C=0)$, $\text{I}(t/\tau)$ decays exponentially as shown in Fig. (\ref{fig:total_dissipation} (a)) and Fig. (\ref{fig:total_dissipation}(b)). However in the long time limit, $\text{I}(t/\tau, C \neq 0)$   decays exponentially as the system eventually reaches steady state. The peak in Fig. (\ref{fig:total_dissipation}(a)) occurs at $t=\frac{\ln \left(\frac{k(C\tau_A+k_B T (\gamma-k\tau_A))}{C\gamma}\right)}{\left(\frac{k}{\gamma}-\frac{1}{\tau_A}\right)}$. With higher $C$ (or $\tau_A$), peak position shifts to higher value of time. On increasing the temperature, the peak in $\text{I}(t/\tau)$ vanishes and $\text{I}(t/\tau)$ shows monotonic decay.

\section{Conclusion}  \label{conclusion}

\noindent In this paper, we analyze the effect of active noise on the dynamics of a single particle where the active noise is modeled with an exponentially correlated non-equilibrium force. We put forward an exact Smoluchowski equation for a single particle subjected to this Gaussian active noise. Our analysis shows that the MSD is supperdiffusive with an exponent $\alpha=2$ in the short time limit and we compare this with an Ornstein-Uhlenbeck process for active noise where the strength of the active noise has an inverse dependence on active correlation time. We believe that such superdiffusive behavior emerges primarily due to the absence of any  FDT for the active noise or in other words, due to persistence motion of active particles, that causes the colloidal particle to have directed motion.
\\
\\
\noindent Another important outcome of our model is that the mean rate of  energy dissipation is non-zero due to the presence of active noise. The force-position correlation in an active system shows an initial growth absent in the case of a purely thermal system, reflecting the fact that there is no FDT for our model of active noise and the detailed balance is broken.
\\
\\
In order to account for non-equilibrium fluctuations, we introduce $T_{eff} (\omega)$ as the ratio between correlation and response of the probe particle.  The frequency dependence of effective temperature is a consequence of breakdown of FDT. In our model system, the response is independent of active noise since both the bath properties are unaffected by the external force. However, in the presence of an external force that perturbs the dynamics of the active particle, the response  depends on the form of the active noise \cite{caprini2018linear}. By adopting a time dependent inverse temperature $T_{\text{eff}}^{-1}(t)$ (inverse Fourier transformation of $\frac{1}{T_{eff}(\omega)}$)  as the replacement for $\delta(t)T^{-1}$, we investigate entropy production in active systems and find that mean rate of entropy production in an active bath is time dependent. A consequence of the breaking of time reversal symmetry in our model is reflected in the time dependence of mean rate of entropy production.   In single molecule experiments, one tracks only a few of the many possible degrees of freedom e.g. the position of the colloidal particle and the orientation of  active particles. Thus we believe that a properly designed experiment should be able to verify our results in future.
\\
\\
It should be noted that in Eq. (\ref{eq:active_noise_correlation}), the activity $C$ is independent of the bacterial correlation time $\tau_A$. In our model both $C$ and $\tau_A$ have the same effect on superdiffusion, dissipation and entropy production. But, if the active noise, $\eta_A(t)$ is drawn from an Ornstein-Uhlenbeck process (OUP), $C$ will be modifed to $\frac{C_0}{\tau_A}$  (Eq. (\ref{eq:ou_correlation})) and in that case, increasing $C_0$ would have the same effect as decreasing $\tau_A$ on superdiffusion, dissipation and entropy production (not shown). A similar trend of the noise strength and persistence time on the dynamics of active glass has been observed by Nandi $et\,\,al.$ \cite{nandi2018random}. They showed if active noise is drawn from an OUP, persistence ($\tau_p$ in their model) promotes glassiness. However, for an exponentially correlated noise that is not drawn from an OUP (as done in the present study), an increase in the persistence time promotes fluidization. However, the choice of the prefactor in the active noise correlation, $C$ or $\frac{C_0}{\tau_A}$ will depend on the microscopic details of the system.
\\
\\
\noindent Our assumption of Gaussianity of active noise is valid when the concentration of active particles in the medium is low and they are weakly interacting. In support, we refer to the work by Solon $et. al.$  where they showed that dilute active systems follow ideal gas like behaviour where the thermal temperature is modified to an active temperature \cite{solon2015pressure}. For dense active systems, active particles have strong interactions and departure from Gaussianity is expected \cite{chaki2019enhanced}.

\section{Acknowledgments} 
\noindent SC acknowledges DST-Inspire for the fellowship.  RC acknowledges SERB for financial support (Project No. SB/SI/PC-55/2013). 

\section{Appendix}
\subsection{Derivation of ASE} \label{appendix_fokker_active}

\noindent Taking averages on both sides of the continuity equation, $\frac{\partial \rho(x,t)}{\partial t}=-\frac{\partial J(x,t)}{\partial x}$

\begin{equation}
\begin{split}
\left<\frac{\partial \rho(x,t)}{\partial t}\right> \equiv \frac{\partial p(x,t)}{\partial t}&=-\frac{1}{\gamma}\frac{\partial}{\partial x} \left<\left[-kx + \xi_T(t) + \eta_A (t) \right]\delta(x(t)-x)\right>\\
&=\frac{k}{\gamma}\frac{\partial}{\partial x}x \left<\delta(x(t)-x)\right>-\frac{1}{\gamma}\frac{\partial}{\partial x} \left< \xi_T(t)\delta(x(t)-x)\right>-\frac{1}{\gamma}\frac{\partial}{\partial x} \left< \eta_A(t)\delta(x(t)-x)\right>\\
&=\frac{k}{\gamma}\frac{\partial}{\partial x}xp(x,t)-\frac{1}{\gamma}\frac{\partial}{\partial x} \left< \xi_T(t)\delta(x(t)-x)\right>-\frac{1}{\gamma}\frac{\partial}{\partial x} \left< \eta_A(t)\delta(x(t)-x)\right>\\
\label{eq:fokker_novikov}
\end{split}
\end{equation}

\noindent To evaluate $\left< \xi_T(t)\delta(x(t)-x)\right>$, we invoke Novikov’s theorem which is applicable to Gaussian random processes \cite{san1980colored,novikov1965functionals}
\begin{equation}
\begin{split}
\left< \xi_T(t)\delta(x(t)-x)\right>&=\int_0^t dt^\prime \left< \xi_T(t) \xi_T(t^{\prime}) \right> \left<\frac{\delta \left[\delta(x(t)-x)\right]}{\delta \xi_T(t^\prime)}\right>\\
&=\int_0^t dt^\prime \left< \xi_T(t) \xi_T(t^{\prime}) \right> \left<-\frac{\partial}{\partial x}\delta(x(t)-x)\frac{\delta x}{\delta \xi_T(t^\prime)}\right>\\
&=-\frac{\partial}{\partial x}\int_0^t dt^\prime \left< \xi_T(t) \xi_T(t^{\prime}) \right> \left<\delta(x(t)-x)\frac{\delta x}{\delta \xi_T(t^\prime)}\right>\\
&=-\frac{\partial}{\partial x}\int_0^t dt^\prime 2 \gamma k_B T \delta(t-t^\prime) \left<\delta(x(t)-x)\frac{\delta x}{\delta \xi_T(t^\prime)}\right>\\
\end{split}
\label{eq:thermal_novikov}
\end{equation}

\noindent Functional differentiation, $\frac{\delta x}{\delta \xi_T(t^\prime)}$ can be easily evaluated from expression  (\ref{eq:langevinact_solution}) \cite{san1980colored},
\begin{equation}
\frac{\delta x(t)}{\delta \xi_T(t^\prime)}=\frac{1}{\gamma} e^{-\frac{k}{\gamma} (t-t^\prime)}
\label{eq:functional_differentiation}
\end{equation}

\begin{equation}
\begin{split}
\textrm{Thus}\,\,\,\,\left< \xi_T(t)\delta(x(t)-x)\right>&=- \frac{2 \gamma k_B T }{\gamma}\frac{\partial}{\partial x}\int_0^t dt^\prime e^{-\frac{k}{\gamma} (t-t^\prime)} \delta(t-t^\prime) \left<\delta(x(t)-x)\right>\\
&=-k_B T \frac{\partial}{\partial x} \left<\delta(x(t)-x)\right>\\
&=-k_B T \frac{\partial}{\partial x}p(x,t)
\end{split}
\label{eq:thermal_final}
\end{equation}

\noindent Since the active noise is also Gaussian, one can proceed as Eq. (\ref{eq:thermal_novikov})

\begin{equation}
\begin{split}
\left< \eta_A(t)\delta(x(t)-x)\right>&=\int_0^t dt^\prime \left< \eta_A(t) \eta_A(t^{\prime}) \right> \left<\frac{\delta \left[\delta(x(t)-x)\right]}{\delta \eta_A(t^\prime)}\right>\\
&=\int_0^t dt^\prime \left< \eta_A(t) \eta_A(t^{\prime}) \right> \left<-\frac{\partial}{\partial x}\delta(x(t)-x)\frac{\delta x}{\delta \eta_A(t^\prime)}\right>\\
&=-\frac{\partial}{\partial x}\int_0^t dt^\prime \left< \eta_A(t) \eta_A(t^{\prime}) \right> \left<\delta(x(t)-x)\frac{\delta x}{\delta \eta_A(t^\prime)}\right>\\
&=-\frac{\partial}{\partial x}\int_0^t dt^\prime Ce^{-\frac{|t-t^{\prime}|}{\tau_A}}\frac{1}{\gamma} e^{-\frac{k}{\gamma} (t-t^\prime)} \left<\delta(x(t)-x)\right>\\
&=-\frac{C}{\gamma}e^{-\left(\frac{k}{\gamma}+\frac{1}{\tau_A}\right)t} \int_0^t dt^\prime e^{\left(\frac{k}{\gamma}+\frac{1}{\tau_A}\right)t^\prime} \frac{\partial}{\partial x} \left<\delta(x(t)-x)\right>\\
&=-\frac{C}{\gamma \left(\frac{k}{\gamma}+\frac{1}{\tau_A}\right)} e^{-\left(\frac{k}{\gamma}+\frac{1}{\tau_A}\right)t} \left(e^{\left(\frac{k}{\gamma}+\frac{1}{\tau_A}\right)t}-1\right) \frac{\partial}{\partial x} \left<\delta(x(t)-x)\right>\\
&=-\frac{C  \left(1-e^{-\left(\frac{k}{\gamma}+\frac{1}{\tau_A}\right)t}\right)}{\gamma \left(\frac{k}{\gamma}+\frac{1}{\tau_A}\right)}  \frac{\partial}{\partial x} p(x,t)\\
&=-A(t) \frac{\partial}{\partial x} p(x,t)
\end{split}
\label{eq:active_novikov}
\end{equation}

\noindent where $A(t)=\frac{C  \left(1-e^{-\left(\frac{k}{\gamma}+\frac{1}{\tau_A}\right)t}\right)}{\gamma \left(\frac{k}{\gamma}+\frac{1}{\tau_A}\right)}$

\noindent After the substitution of Eq.  (\ref{eq:thermal_final}) and  (\ref{eq:active_novikov}) into Eq.  (\ref{eq:fokker_novikov}),

\begin{equation}
\begin{split}
 \frac{\partial p(x,t)}{\partial t}&=\frac{k}{\gamma}\frac{\partial}{\partial x}xp(x,t)+\left(\frac{k_B T}{\gamma}+\frac{A(t)}{\gamma}\right) \frac{\partial^2}{\partial x^2}p(x,t)\\
\label{eq:active_fokker_appendix}
\end{split}
\end{equation}

\subsection{Calculation of MSD} \label{MSD_appendix}

\begin{equation}
\begin{split}
 \frac{\partial }{\partial t}\left<x^2(t)\right>\equiv  \frac{\partial }{\partial t}\int_{-\infty}^{\infty}dx x^2 p(x,t)&=\frac{k}{\gamma}\int_{-\infty}^{\infty}dx x^2\frac{\partial}{\partial x}xp(x,t)+\left(D+D_A(t)\right)\int_{-\infty}^{\infty}dx x^2\frac{\partial^2}{\partial x^2}p(x,t)\\
&=\frac{k}{\gamma}\left[x^3 p(x,t)\right]_{-\infty}^{\infty} - \frac{k}{\gamma} \int_{-\infty}^{\infty}dx 2 x^2 p(x,t) \\
&+ \left(D+D_A(t)\right) \left[x^2 \frac{\partial}{\partial x}p(x,t)\right]_{-\infty}^{\infty} -\left(D+D_A(t)\right) \int_{-\infty}^{\infty}dx 2x\frac{\partial}{\partial x }p(x,t)\\
&=-\frac{2k}{\gamma}\left<x^2(t)\right>-2\left(D+D_A(t)\right) \left[x p(x,t)\right]_{-\infty}^{\infty}\\
&+2\left(D+D_A(t)\right) \int_{-\infty}^{\infty}dx p(x,t)\\
&=-\frac{2k}{\gamma}\left<x^2(t)\right>-2\left(D+D_A(t)\right) \left[x p(x,t)\right]_{-\infty}^{\infty}\\
&=-\frac{2k}{\gamma}\left<x^2(t)\right>+2\left(D+D_A(t)\right)
\label{eq:active_mean_square_equation}
\end{split}
\end{equation}

\noindent Multiplying both sides of Eq.  (\ref{eq:active_mean_square_equation}) by  integrating factor, $e^{\frac{2k}{\gamma}t}$ , integrating  and putting proper limits, 

\begin{equation}
\begin{split}
e^{\frac{2k}{\gamma}t}\left<x^2(t)\right>&=2D\int_{-\infty}^{t}dt^{\prime}e^{\frac{2k}{\gamma}t}+2\int_{0}^{t}dt^{\prime}e^{\frac{2k}{\gamma}t}D_A(t)\\
&=\frac{2D\gamma}{2k}e^{\frac{2k}{\gamma}t}+\frac{2 C }{\gamma^2 \left(\frac{k}{\gamma}+\frac{1}{\tau_A}\right)}\int_{0}^{t}e^{\frac{2k}{\gamma}t} \left(1-e^{-\left(\frac{k}{\gamma}+\frac{1}{\tau_A}\right)t}\right)\\
&=\frac{k_B T}{k}e^{\frac{2k}{\gamma}t}+\frac{k_B T_A}{k}\left(e^{\frac{2k}{\gamma}t}-1\right)-\frac{2C\left(e^{\left(\frac{k}{\gamma}-\frac{1}{\tau_A}\right)t}-1\right)}{\gamma^2\left(\frac{k^2}{\gamma^2}-\frac{1}{\tau_A^2}\right)} \\
\label{eq:active_mean_square_fluctuation}
\end{split}
\end{equation}

\noindent Thus 
\begin{equation}
\begin{split}
\left<x^2(t)\right>&=\frac{k_B T}{k}+\frac{k_B T_A}{k}\left(1-e^{-\frac{2k}{\gamma}t}\right)-\frac{2C\left(e^{-\left(\frac{k}{\gamma}+\frac{1}{\tau_A}\right)t}-e^{-\frac{2k}{\gamma}t}\right)}{\gamma^2\left(\frac{k^2}{\gamma^2}-\frac{1}{\tau_A^2}\right)} 
\label{eq:active_mean_square}
\end{split}
\end{equation}

\begin{equation}
\begin{split}
\left<\left(x(t)-x(0)\right)^2\right>&=\left<x_0^2\right>\left(e^{-\frac{k}{\gamma}t}-1\right)^2+\frac{e^{-\frac{2k}{\gamma}t}}{\gamma^2}\int_0^t dt^\prime\int_0^{t} dt^{\prime\prime} e^{\frac{k}{\gamma} (t^\prime+t^{\prime\prime})}\left(\left<\xi (t^\prime)\xi (t^{\prime\prime})\right>+\left<\eta_A (t^\prime)\eta_A (t^{\prime\prime})\right>\right)\\
&=\left<x_0^2\right>\left(e^{-\frac{k}{\gamma}t}-1\right)^2+\frac{2k_B T\gamma e^{-\frac{2k}{\gamma}t}}{\gamma^2}\int_0^t dt^\prime\int_0^{t} dt^{\prime\prime} e^{\frac{k}{\gamma} (t^\prime+t^{\prime\prime})}\delta(t^\prime-t^{\prime\prime})\\
&+\frac{C e^{-\frac{2k}{\gamma}t}}{\gamma^2}\int_0^t dt^\prime\int_0^{t} dt^{\prime\prime} e^{\frac{k}{\gamma} (t^\prime+t^{\prime\prime})}e^{-\frac{|t^\prime-t^{\prime\prime}|}{\tau_A}}\\
&=\frac{k_B T}{k}\left(e^{-\frac{2k}{\gamma}t}+1-2e^{-\frac{k}{\gamma}t}\right)+\frac{2k_B T e^{-\frac{2k}{\gamma}t}}{\gamma}\int_0^t dt^\prime e^{\frac{2k}{\gamma} t^\prime}\\
&+\frac{2C e^{-\frac{2k}{\gamma}t}}{\gamma^2}\int_0^t dt^\prime\int_0^{t^\prime} dt^{\prime\prime} e^{\frac{k}{\gamma} (t^\prime+t^{\prime\prime})}e^{-\frac{(t^\prime-t^{\prime\prime})}{\tau_A}}\\
&=\frac{2k_B T}{k}\left(1- e^{-\frac{k}{\gamma} t}\right)+\frac{C }{k\gamma (\frac{k}{\gamma}+\frac{1}{\tau_A})}\left(1-e^{-\frac{2k}{\gamma}t}\right)-\frac{2C }{(\frac{k^2}{\gamma^2}-\frac{1}{\tau_A^2})\gamma^2}\left(e^{-(\frac{k}{\gamma}+\frac{1}{\tau_A})t}-e^{-\frac{2k}{\gamma}t}\right)
\label{eq:msd_act}
\end{split}
\end{equation}

\subsection{Short time superdiffusion} \label{superdiffusion_appendix}

\begin{equation}
\begin{split}
\left<\left(x(t)-x_0\right)^2\right>&\approx \frac{2k_BT}{k}\left[1-\left(1-\frac{k t}{\gamma}\right)\right]+\frac{C}{k\gamma\left(\frac{k}{\gamma}+\frac{1}{\tau_A}\right)}\left[1-\left(1-\frac{2k t}{\gamma}\right)\right]\\
&-\frac{2C }{(\frac{k^2}{\gamma}-\frac{1}{\tau_A^2})\gamma^2}\left(e^{-(\frac{k}{\gamma}+\frac{1}{\tau_A})t}-\left(1-\frac{2k}{\gamma}t\right)\right)\\
&=\frac{2k_BTt}{\gamma}+\frac{C\tau_A}{k\gamma}\frac{2k t}{\gamma}+\frac{2C \tau_A^2}{\gamma^2}\left(e^{-\frac{1}{\tau_A}t}-\left(1-\frac{2k}{\gamma}t\right)\right)\\
&=\frac{2k_BTt}{\gamma}+\frac{2C\tau_At}{\gamma^2}+\frac{2C \tau_A^2}{\gamma^2}\left(1-\frac{ t}{\tau_A}+\frac{t^2}{2\tau_A^2}-\left(1-\frac{2k}{\gamma}t\right)\right)\\
&=\frac{2k_BTt}{\gamma}+\frac{2C\tau_At}{\gamma^2}+\frac{2C \tau_A^2}{\gamma^2}\left(-\frac{ t}{\tau_A}+\frac{t^2}{2\tau_A^2}+\frac{2k}{\gamma}t\right)\\
&=\frac{2k_BTt}{\gamma}+\frac{2C\tau_At}{\gamma^2}+\frac{2C \tau_A^2}{\gamma^2}\left(-\frac{ t}{\tau_A}+\frac{t^2}{2\tau_A^2}\right)\\
&=\frac{2k_BTt}{\gamma}+\frac{C t^2}{\gamma^2}\\
&=2Dt+\frac{C t^2}{\gamma^2}
\label{eq:short}
\end{split}
\end{equation}

\subsection{Force-position correlation for active noise} \label{active_dissipation_appendix}

\begin{equation}
\begin{split}
\left< x(t)\eta_A(0) \right> & =\left<x_0 \eta_A(0) \right>e^{-\frac{k}{\gamma}t}+\frac{1}{\gamma}\int_0^t dt^\prime e^{-\frac{k}{\gamma} (t-t^\prime)}\left[\left<\xi_T(t^\prime)\eta_A(0)\right>+\left<\eta_A (t^\prime)\eta_A(0)\right>\right]\\
&=\frac{e^{-\frac{k}{\gamma}t}}{\gamma}\int_0^t dt^\prime e^{\frac{k}{\gamma} t^\prime}\left<\eta_A (t^\prime)\eta_A(0)\right>\\
&=\frac{Ce^{-\frac{k}{\gamma}t}}{\gamma}\int_0^t dt^\prime e^{\frac{k}{\gamma} t^\prime}e^{-\frac{ t^\prime}{\tau_A}}\\
&=\frac{C\left[e^{-\frac{t}{\tau_A}}-e^{-\frac{k}{\gamma}t}\right]}{\gamma\left(\frac{k}{\gamma}-\frac{1}{\tau_A}\right)}
\label{eq:active_pos_force_correlation_appendix}
\end{split}
\end{equation}

\subsection{Force-position correlation for thermal noise} \label{thermal_dissipation_appendix}
\begin{equation}
\begin{split}
\left< x(t)\xi_T(0) \right> & =\left<x_0 \xi_T(0) \right>e^{-\frac{k}{\gamma}t}+\frac{1}{\gamma}\int_0^t dt^\prime e^{-\frac{k}{\gamma} (t-t^\prime)}\left[\left<\xi_T(t^\prime)\xi_T(0)\right>+\left<\eta_A (t^\prime)\xi_T(0)\right>\right]\\
&=\frac{e^{-\frac{k}{\gamma}t}}{\gamma}\int_0^t dt^\prime e^{\frac{k}{\gamma} t^\prime}\left<\xi_T(t^\prime)\xi_T(0)\right>\\
&=\frac{2k_B T \gamma e^{-\frac{k}{\gamma}t}}{\gamma}\int_0^t dt^\prime e^{\frac{k}{\gamma} t^\prime} \delta(t^\prime)\\
&=k_B T  e^{-\frac{k}{\gamma}t}
\label{eq:thermal_pos_force_correlation_appendix}
\end{split}
\end{equation}

\bibliographystyle{apsrev}


\end{document}